\documentclass[structabstract]{aa}
\usepackage{amsmath}

\usepackage{graphicx}
\usepackage{txfonts}
\usepackage{epsfig}

\begin{document}

   \title{Super-Eddington wind scenario for the progenitors of type Ia supernovae:
Accreting He-rich matter onto white dwarfs}

   \author{B. Wang \inst{1,2}
          \and
          Y. Li \inst{1,2}
          \and
          X. Ma \inst{1,2}
          \and
          D.-D. Liu \inst{1,2}
          \and
          X. Cui \inst{1,2}
          \and
          Z. Han \inst{1,2}
          }

   \institute{Yunnan Observatories, Chinese Academy of Sciences, Kunming 650216, China;
              \email{wangbo@ynao.ac.cn}
              \and
              Key Laboratory for the Structure and Evolution of Celestial Objects, Chinese Academy of Sciences, Kunming 650216, China
              }
   \date{Received; accepted}

\abstract
{Supernovae of type Ia  (SNe Ia) are believed to be thermonuclear
explosions of carbon-oxygen white dwarfs (CO WDs). However, the mass
accretion process onto CO WDs is still not completely understood.}
{In this paper, we  study the accretion of He-rich matter onto CO WDs and explore a scenario
in which a strong wind forms on the surface of the WD if the total luminosity exceeds the Eddington limit.}
{Using a stellar evolution code called modules for
experiments in stellar astrophysics (MESA), we simulated the
He accretion process onto CO WDs for WDs with masses of
0.6$-$$1.35\,{M}_\odot$ and various accretion rates of $10^{-8}$$-$$10^{-5}\,M_\odot\,\mbox{yr}^{-1}$.}
{If the contribution of the total luminosity is included when determining the Eddington accretion rate, then a
super-Eddington wind could be triggered at relatively lower accretion rates than those of previous studies based on steady-state models.
The super-Eddington wind can prevent the WDs with high accretion rates from evolving into red-giant-like He stars.
We found that the contributions from thermal energy of the WD
are non-negligible, judging by our simulations, even though the nuclear burning energy is the
dominating source of luminosity.
We also provide the limits of the steady He-burning regime in
which the WDs do not lose any accreted matter and increase their mass steadily,
and calculated the mass retention efficiency during He layer flashes
for various WD masses and accretion rates. These obtained results
can be used in future binary population synthesis computations.}
{}


\titlerunning{Accreting He-rich matter onto WDs}

\authorrunning{B. Wang et al.}

   \maketitle


\section{Introduction}

Supernovae of type Ia (SNe Ia) are of great importance in astrophysics, especially in Cosmology
and the chemical evolution of galaxies  (e.g., Howell 2011; Matteucci \& Greggio 1986).
They are also accelerators of cosmic rays (see Fang \& Zhang 2012).
These supernovae are thought to originate from thermonuclear explosions of carbon-oxygen white
dwarfs (CO WDs) in binaries, although the mass donor is still under debate (see Podsiadlowski et al. 2008).
The mass donor could be a non-degenerate star in the single degenerate model
(e.g., Whelan \& Iben 1973; Nomoto et al. 1984; Hachisu et al. 1996; Li \& van den Heuvel 1997;
Han \& Podsiadlowski 2004, 2006; Wang et al. 2010; Meng \& Podsiadlowski 2013, 2014) or
another WD in the double degenerate model (e.g., Iben \& Tutukov 1984; Webbink 1984; Chen et al. 2012).
To date, two explosion models of SNe Ia have been discussed frequently,
the Chandrasekhar mass model and the sub-Chandrasekhar mass model.
Recent reviews of the theoretical models of SNe Ia and the observational constraints include
Wang \& Han (2012), Hoeflich et al. (2013), Hillebrandt et al. (2013), Wang et al. (2013a) and Maoz et al. (2014).

The He accretion process onto WDs plays a critical role in
understanding the progenitor models of SNe Ia. First, in the classical Chandrasekhar mass model, a WD could directly accrete matter
from a non-degenerate or degenerate He-rich companion. The accreted He-rich matter
is burned into carbon and oxygen, leading to the mass increase of the WD. The WD
may explode as an SN Ia when it grows in mass to the Chandrasekhar limit
(e.g., Yoon \& Langer 2003; Wang et al. 2009a; Piersanti et al. 2014).
Second, in the sub-Chandrasekhar mass model, the explosion of a CO WD may be triggered by the detonation of
a substantial surface layer of accreted He, where the mass donor could be a non-degenerate He star or
a He WD. This model is also known as the He double-detonation model
(see, e.g., Woosley et al.\ 1986; Livne 1990; Fink et al. 2007; Wang et al. 2013b; Piersanti et al. 2014).

The He accretion process onto WDs is also related to the
formation of some peculiar objects, such as
He novae (e.g., Kato et al. 2000; Ashok \& Banerjee 2003)
and AM CVn systems (e.g., Nelemans et al. 2001; Shen \& Bildsten 2009;
Brooks et al. 2015; Piersanti et al. 2015). Meanwhile,
single He layer detonation at the surface of accreting
WDs may produce some faint and fast transients like the hypothetical .Ia SNe (see, e.g., Bildsten
et al. 2007; Shen et al. 2010;  Woosley \& Kasen 2011;
Sim et al. 2012; Kasliwal et al. 2012; Brooks et al. 2015).
In addition, Wang et al. (2013b) recently argued that
some type Iax SNe (a kind of sub-luminous SNe Ia) may be explained by the specific class of He-ignited
WD explosions. Furthermore, many recent binary population synthesis studies
also involved the He accretion onto the surface of WDs
(see, e.g., Wang et al. 2009b, 2013b; Ruiter et al. 2009, 2013, 2014;
Toonen et al. 2014; Claeys et al. 2014).

In the observations, there are many massive WD+He star binaries,
e.g., CD$-$30$^{\circ}$\,11223, KPD 1930+2752, V445 Puppis, and HD 49798 with its WD companion. 
It has been  suggested that they are all  progenitor candidates of SNe Ia (e.g.,
Geier et al. 2007, 2013; Kato et al. 2008; Woudt et al. 2009;
Mereghetti et al. 2009; Wang \& Han 2010).
In addition, the companion of SN 2012Z was probably a He star (e.g.,
McCully et al. 2014; Wang et al. 2014a; Liu et al. 2015), and
the mass donor of SN 2014J may also be a He star (see Diehl et al. 2014).
Additionally, US 708 (a hypervelocity He star) is likely to be the ejected donor remnant
of an SN Ia that originated from a WD binary where the mass donor  is a He star
(e.g., Geier et al. 2015; see also Wang \& Han 2009; Justham et al. 2009).

However, the He accretion process onto CO WDs is still not clearly understood.
If the accretion rate is too low, a He layer flash happens on the surface of the WD (e.g., Kato \& Hachisu 2004).
If the accretion rate of the WD is too high,  it will become a red-giant-like He star
owing to the accumulation of the accreted matter on its surface (e.g., Nomoto 1982).
Therefore, the steady He layer burning only occurs in a narrow regime, in which the accreted
He can be completely burned into carbon and oxygen.
Wang et al. (2009a) recently investigated the WD+He star channel of  SN Ia progenitors
on the basis of an optically thick wind assumption (e.g., Hachisu et al. 1996). In this assumption,
the red-giant-like regime was replaced by the optically
thick wind regime; the He-rich matter can be
transformed into carbon and oxygen at a critical rate in this regime, whereas
the unprocessed matter is blown away via the optically thick wind.
However, this assumption is still  under debate
(e.g., Langer et al. 2000), and the metallicity threshold predicted by this model
has not been observed (e.g., Badenes et al. 2009).

At low accretion rates, the accreting WDs will undergo multicycle
He layer flashes like nova outbursts, but previous studies
mainly involved one cycle when simulating He layer flashes (e.g., Kato \& Hachisu 2004).
In this paper, we also investigate a multicycle evolution of
the He layer flash with various WD masses and
obtain the mass retention efficiency during He layer flashes.

The purpose of this paper is to study the He layer
burning on the surface of WDs. These WDs have masses from 0.6 to
$1.35\,{M}_\odot$ with various accretion rates.
In Sect. 2, the
basic assumptions and methods for numerical calculations are described.
Our numerical calculation results are given  in Sect. 3.
Finally, we present the discussion and summary in Sect. 4.

\section{Numerical code and methods}

Employing the stellar evolution code called modules for
experiments in stellar astrophysics (MESA, version 3661; see Paxton et al. 2011, 2013),
we simulated the accretion process of He-rich matter onto CO WDs.
The MESA default OPAL opacity is adopted in our simulations (see Fig. 2 in Paxton et al. 2011).
Twenty-one isotopes  are included in the nuclear network, which are coupled by 50 nuclear reactions.
We selected two MESA suite cases for our numerical calculations:  \texttt{make\_co\_wd}
and \texttt{wd2}.

We used the MESA suite case \texttt{make\_co\_wd} to construct CO WD
models. In order to get the initial CO WD models, we chose zero age main-sequence stars with
intermediate masses and evolved them to the cooling sequence. We obtained a series of CO WD models with
masses from $0.6$ to $1.0\,{M}_\odot$ based on this method.
To obtain CO WDs with masses $>$$1.0\,{M}_\odot$,
we added the mass to our $1.0\,{M}_\odot$ WD model
by accreting matter with its surface composition.
Finally, we constructed ten hot WD models
as our initial models with masses in the range of $0.6$--$1.35\,{M}_\odot$.
We note that we came across some numerical problems when simulating the evolution of cold
WDs during mass accretion, possibly owing to the high degeneracy of matters in their
outer layers.\footnote{At the beginning of the He layer burning,
the evolution of cold WDs may
be somewhat different to that of the hot ones during mass accretion.
However, the difference is not significant
once  an equilibrium condition is reached.
Iben (1982) also used hot WD models
when simulating the evolution of accreting
WDs (see also  Iben \& Tutukov 1989).}
Thus, we selected hot WD models at the top
of the WD cooling track as our initial models in the
simulations of mass accretion.
The ten WD models have masses (central temperatures) of 0.6$\,{M}_\odot$ ($7.1\times10^7{\rm K}$), 0.7$\,{M}_\odot$ ($7.0\times10^7{\rm K}$) , 0.8$\,{M}_\odot$ ($6.9\times10^7{\rm K}$), 0.9$\,{M}_\odot$ ($7.7\times10^7{\rm K}$), 1.0$\,{M}_\odot$ ($10.0\times10^7{\rm K}$), 1.1$\,{M}_\odot$ ($10.4\times10^7{\rm K}$), 1.2$\,{M}_\odot$ ($12.2\times10^7{\rm K}$), 1.25$\,{M}_\odot$ ($13.7\times10^7{\rm K}$),
1.3$\,{M}_\odot$ ($16.0\times10^7{\rm K}$), and 1.35$\,{M}_\odot$ ($20.7\times10^7{\rm K}$).\footnote{The temperatures adopted here are somewhat high, which may lead to
a high thermal radiation contribution from the CO core
for an old WD to the luminosity.}
The metallicity of these WDs is set to be 0.02.

We used the MESA suite case \texttt{wd2} to calculate the He accretion
onto the surface of CO WDs. The suite case \texttt{wd2} controls mass accretion and includes
an acceleration term in the equation of hydrostatic equilibrium so that we can
calculate the mass ejection processes during nova outbursts.
The accreted matter is composed of He mass fraction $Y=0.98$, and metallicity
$Z=0.02$. We carried out a series of calculations for all of our  WD models
with various accretion rate of $10^{-8}$$-$$10^{-5}\,M_\odot\,\mbox{yr}^{-1}$.
In each simulation, we calculated for a long time to get detailed information about accretion.

The luminosity $L$ of the
accreting WD may exceed the Eddington limit once the accretion rate is too high.
In this case, a super-Eddington wind can be triggered.
The Eddington luminosity can be expressed as
\begin{equation}
{L}_{\rm Edd}=\frac{4\pi cGM_{\rm WD}}{\kappa},
\end{equation}
in which $M_{\rm WD}$, $c$, $G$, and $\kappa$  are the WD mass, vacuum
speed of light, the gravitational constant, and the
opacity, respectively.

The luminosity ${L}$ of accreting WDs usually has three contributors:
nuclear burning energy, thermal energy,
and the accretion energy released by the accreted matter.
Among these three contributors,
the nuclear burning luminosity accounts for most of the total luminosity of the WD.
The values of the nuclear burning luminosity are always much higher than the other two contributors.
The nuclear burning luminosity ${L}_{\rm nuc}$ can be expressed as
\begin{equation}
L_{\rm nuc}=YQ\dot{M}_{\rm nuc},
\end{equation}
in which $\dot{M}_{\rm nuc}$, $Q$, and $Y$ are
the nuclear burning rate of the accreted He, the nuclear energy released by
per unit mass of He, and
the He mass fraction in the accreted matter,
respectively.
The accretion luminosity ${L}_{\rm acc}$ due to the gravitational energy released is
\begin{equation}
{L}_{\rm acc}=\frac{{GM_{\rm WD}}\dot{M}_{\rm acc}}{R_{\rm WD}},
\end{equation}
in which $\dot{M}_{\rm acc}$ and $R_{\rm WD}$ are the accretion rate
and the WD radius, respectively.
The accretion energy deposited at or near the
photosphere of the WD  does not
get absorbed by the WD, as heat always escapes
much more quickly  than the speed at which the fluid moves
inward  (e.g., Townsley \& Bildsten 2004).
Because the accretion energy  can be
radiated away from the WD surface rapidly,
here we define the luminosity $L_{\rm prime}$ as
$(L-{L}_{\rm acc})$   (see also Ma et al. 2013).

By setting $L_{\rm acc}=L_{\rm Edd}$, the Eddington accretion rate can be expressed as
\begin{equation}
\dot{M}_{\rm Edd}=4\pi cR_{\rm WD}/\kappa,
\end{equation}
which was adopted in Nomoto (1982).
In this paper, we first set ${L}_{\rm acc}={L}_{\rm Edd}$  as
the wind triggering criteria to reproduce the results of
the steady-state models  of Nomoto (1982),
which can be used to test the reliability of our code.
Then we set $L_{\rm prime}=L_{\rm Edd}$ as the wind triggering criteria to investigate the
long-term evolution of WDs during He accretion.
 In the steady
He burning regime, we note that $L$ and $L_{\rm prime}$ are not so different from
each other because $L_{\rm nuc}$ is far greater than $L_{\rm acc}$.
Thus, the distinction between using $L$ or $L_{\rm prime}$ as the wind triggering criterion
is relatively small.
Additionally, if $L_{\rm nuc}\,=\,L_{\rm Edd}$ is assumed and the accreted matter is burned completely
(i.e., $\dot{M}_{\rm acc} =  \dot{M}_{\rm nuc}$), then the Eddington accretion rate can be expressed as
$\dot{M}_{\rm Edd}=4\pi cGM_{\rm WD}/(YQ\kappa)$ (see also Shen \& Bildsten 2007).

\section{Numerical results}
\subsection{Reproducing previous results}

\begin{figure}[]
\begin{center}
\includegraphics[width=8.5cm,angle=0]{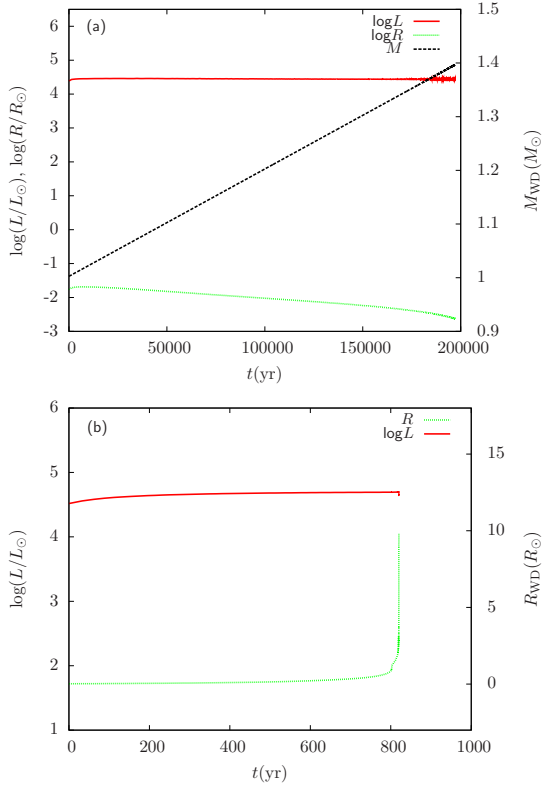}
 \caption{The evolution of the luminosity, radius, and mass of an accreting WD with $1\,{M}_\odot$
in the steady He layer burning regime ($\dot{M}_{\rm acc}=2\times 10^{-6}\,{M}_\odot\,\mbox{yr}^{-1}$; panel a),
and in the red-giant-like He star regime ($\dot{M}_{\rm acc}=4\times10^{-6}\,{M}_\odot\,\mbox{yr}^{-1}$; panel b).}
  \end{center}
\end{figure}

In order to compare our results with those of the steady-state models of Nomoto (1982), we studied the
evolution of CO WDs during He accretion by setting $L_{\rm acc}=L_{\rm Edd}$.
Two representative examples of our
calculations are shown in Fig. 1. In panel (a), the He layer burning is steady on the surface of the WD
at an accretion rate of $2\times 10^{-6}\,{M}_\odot\,\mbox{yr}^{-1}$. In this case,
the WD can increase its mass to the condition of the explosive carbon ignition, resulting in an
SN Ia explosion as a consequence.
The WD explosion mass can exceed the Chandrasekhar limit if rotation is taken into account 
(e.g., Yoon \& Langer 2004; Chen \& Li 2009; Justham 2011; Hachisu et al. 2012; Wang et al. 2014b).
In panel (b), the WD will evolve into a red-giant-like He star because of a high accretion rate of $4\times
10^{-6}\,{M}_\odot\,\mbox{yr}^{-1}$ that is higher than the critical value of $\dot{M}_{\rm RG}$ presented in Fig. 2.
In such a case, the WD will not form an SN Ia but will evolve into a red-giant-like He star.

\begin{figure}[]
\begin{center}
\includegraphics[width=8.5cm,angle=0]{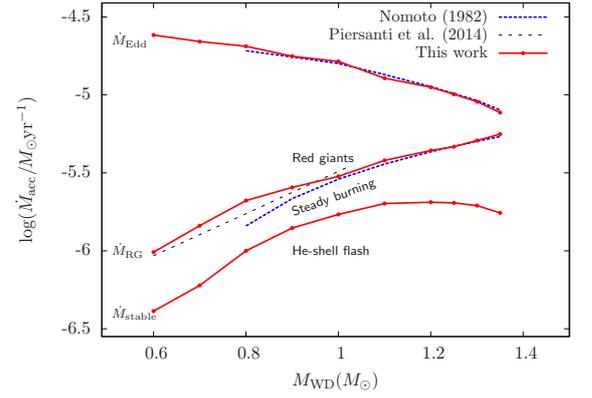}
 \caption{Properties of the He layer burning on the surface of WD in the
plane of WD mass and accretion rate, in which we set $L_{\rm acc}=L_{\rm Edd}$.
The red solid lines are the results of our simulations,
whereas the blue dotted line and the black dashed line are taken from Nomoto (1982) and Piersanti et al. (2014), respectively.
The Eddington accretion rate $\dot{M}_{\rm Edd}$ in this case is much higher than the values of $\dot{M}_{\rm RG}$.
We note that Piersanti et al. (2014) did not present the Eddington accretion rate in their paper.}
  \end{center}
\end{figure}

In Fig. 2, we show the stable He burning regime
by setting $L_{\rm acc}=L_{\rm Edd}$, in which the WD can increase its mass steadily in this regime.
The critical values of $\dot{M}_{\rm RG}$ and the minimum accretion rate $\dot{M}_{\rm st}$ for stable He layer burning  here
are approximated by the  algebraic form
\begin{equation}
\small{\dot{M}_{\rm RG}=2.156\times10^{-6}({M}_{\rm WD}^2+0.820{M}_{\rm WD}-0.379)},
\end{equation}
\begin{equation}
\small{\dot{M}_{\rm st}=1.132\times10^{-5}(-{M}_{\rm WD}^3+
2.562{M}_{\rm WD}^2-1.845{M}_{\rm WD}+0.435)},
\end{equation}
where ${M}_{\rm WD}$ is in units of ${M}_\odot$, and $\dot{M}_{\rm
RG}$ and $\dot{M}_{\rm st}$ are in units of
$M_\odot\,\mbox{yr}^{-1}$. These two critical values strongly depend on the WD mass and
were determined through the bisection method for each WD mass.
Here, we adopt the Thomson opacity for comparison with previous studies.
In this figure, we also show the Eddington accretion rate,
which is very close to that of Nomoto (1982).

The WD will evolve into a red-giant-like He star
if the accretion rate is larger than  $\dot{M}_{\rm RG}$,
above which the envelope of the WD expands to red-giant dimensions and outflow may occur.
In Fig. 2, we also compared the values of $\dot{M}_{\rm RG}$  between this work
and those of the steady-state models of Nomoto (1982) and
Piersanti et al. (2014).
It seems that our results are close to those of  Piersanti et al. (2014).
However, there are still some small differences
with the results of Nomoto (1982) when $M_{\rm WD}<1\,{M}_\odot$, which likely results from the different methods adopted.
Nomoto (1982) studied the stability of the steady-state models via a linear stability analysis,
whereas we performed a series of stellar evolution simulations  by adopting a detailed He accretion
process.

\subsection{Super-Eddington wind scenario}

We calculated the Eddington
accretion rate $\dot{M}_{\rm Edd}$ for various WD masses
using $L_{\rm prime}=L_{\rm Edd}$ as a criterion for
triggering the super-Eddington wind.
The final results are presented in Fig. 3. Here, we used the opacity of the
photosphere when calculating the Eddington luminosity as was done in
Denissenkov et al. (2013) (see also Ma et al. 2013).
The values of the Eddington critical
accretion rate $\dot{M}_{\rm Edd}$ are approximated by the following
algebraic form:
\begin{equation}
\small{\dot{M}_{\rm Edd}=1.702\times10^{-5}({M}_{\rm WD}^3-
3.202{M}_{\rm WD}^2+3.561{M}_{\rm WD}-1.221)},
\end{equation}
where ${M}_{\rm WD}$ is in units of ${M}_\odot$, and $\dot{M}_{\rm
Edd}$ is in units of $M_\odot\,\mbox{yr}^{-1}$.

\begin{figure}[]
\begin{center}
\includegraphics[width=8.5cm,angle=0]{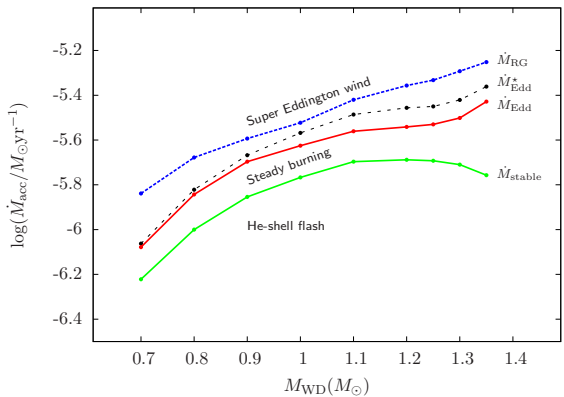}
 \caption{Properties of the He layer burning on the surface of WD in the
plane of WD mass and accretion rate, in which the super-Eddington wind
is triggered when $L_{\rm prime}=L_{\rm Edd}$. The red solid line shows
the Eddington accretion rate $\dot{M}_{\rm Edd}$, whereas the green solid
line presents the minimum accretion rate $\dot{M}_{\rm st}$ for stable He layer burning.
For comparison, we also present the boundaries of
$\dot{M}^{\star}_{\rm Edd}$ (black dashed line) by setting
$L_{\rm nuc}=L_{\rm Edd}$  and $\dot{M}_{\rm RG}$ (blue dotted line).}
  \end{center}
\end{figure}

The values of $\dot{M}_{\rm Edd}$ obtained here are much lower than those in Nomoto (1982) since
$L_{\rm prime}$ is far greater than the accretion luminosity.
We note that the values of $\dot{M}_{\rm Edd}$
are even lower than $\dot{M}_{\rm RG}$, which is caused by a high value of opacity.
For the case of $1\,{M}_\odot$ WD with
an accretion rate of $2.14\times 10^{-6}\,M_\odot\,\mbox{yr}^{-1}$,
$L_{\rm prime}$ can reach the Eddington critical limit.
The temperature at the photosphere for this accreting WD is $9.15\times10^5{\rm K}$, and the density is
$4.07\times10^{-4}{\rm g}/{\rm cm}^{3}$.
The opacity in this case is $0.427\,{\rm cm}^{2}/{\rm g}$, which is
higher than that for electron scattering in fully ionized He ($\kappa\sim0.2\,{\rm cm}^{2}/{\rm g}$).
In addition to the contribution of electron scattering, the free-free transition of electrons also
contributes to the opacity of the photosphere for the accreting WD.

For comparison, we also give the Eddington critical
accretion rate $\dot{M}^{\star}_{\rm Edd}$ by setting $L_{\rm nuc}=L_{\rm Edd}$ as
the wind triggering criteria (see the black dashed line in Fig. 3).
The contributions from thermal energy are non-negligible
in our simulations even though $L_{\rm nuc}$ dominates the luminosity. For hot massive WDs,
the thermal energy can account for up to 20\% of $L_{\rm prime}$.
The values of $\dot{M}^{\star}_{\rm Edd}$ are approximated by the
algebraic form
\begin{equation}
\small{\dot{M}^{\star}_{\rm Edd}=0.953\times10^{-5}({M}_{\rm WD}^3-
3.352{M}_{\rm WD}^2+4.161{M}_{\rm WD}-1.528)},
\end{equation}
where ${M}_{\rm WD}$ is in units of ${M}_\odot$,
and $\dot{M}^{\star}_{\rm Edd}$ is in units of $M_\odot\,\mbox{yr}^{-1}$.

The red-giant regime is now replaced by a super-Eddington wind regime.
In this regime, the super-Eddington wind can be formed, which can blow away the surface matter of the  WD and
prevent the He-rich envelope from expanding. In this case, the WD will
never become a red-giant-like He star.
The He layer burning in this regime is
steady on the  surface of the WD, in which the accreted He-rich matter is burned into carbon and oxygen at a rate of
$\dot{M}_{\rm Edd}$; the unprocessed matter can be blown away by the super-Eddington wind
with a mass-loss rate of $(\dot M_{\rm acc}-\dot{M}_{\rm Edd})$.
Therefore, the super-Eddington wind provides an
alternative way to the optically thick wind, in the sense that it can prevent an accreting
WD from expanding to a red-giant-like He star. In comparison to the
optically thick wind, the super-Eddington wind is not dependent on
the metallicity (see Fig. 4 of Ma et al. 2013).
We note that the accretion rate $\dot M_{\rm acc}$ might have an upper
limit (e.g., 3$\dot{M}_{\rm Edd}$) above which all the unprocessed matter
cannot be blown away by the super-Eddington wind,
resulting in the formation of a common envelope (e.g., Langer et al. 2000; Tauris et al. 2013).

\subsection{Mass retention efficiency in He layer flashes}

\begin{figure}[]
\begin{center}
\includegraphics[width=9.cm,angle=0]{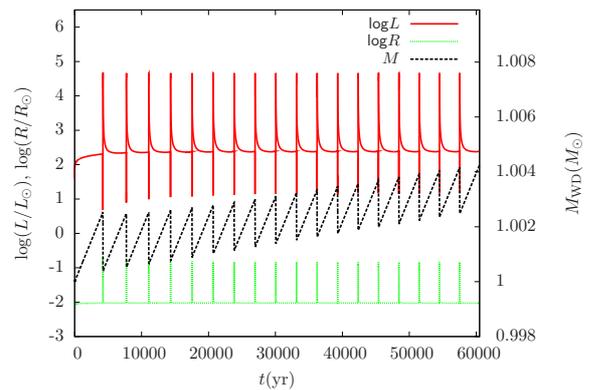}
 \caption{Representative example of He layer flashes on
 the surface of a $1\,{M}_\odot$ WD , in which we set
 $\dot{M}_{\rm acc}=6\times10^{-7}\,{M}_\odot\,\mbox{yr}^{-1}$.}
  \end{center}
\end{figure}

When the accretion rate is smaller than $\dot{M}_{\rm st}$, a He layer flash
occurs on the surface of the WD to develop a nova outburst.
The mass loss via the wind carries away a
part of the envelope matter if the He layer flashes happen,  which reduces the  mass-growth rate of the WD.
According to the stellar evolution code MESA, multi-cycle nova
evolutionary sequences with CO WD cores were recently constructed by Denissenkov et al. (2013),
in which ${L}_{\rm prime}=L_{\rm Edd}$ was employed as the triggering criteria
for the super-Eddington wind. As was done in the works of Denissenkov et al. (2013),
we also assume that the  rate at which the mass-loss
kinetic energy changes is determined by the excess of nova luminosity over the
Eddington luminosity.

We obtained the mass retention efficiency of He layer flashes for various WD masses.
In Fig. 4, we show a representative example of He layer flashes on the surface of a $1\,{M}_\odot$ WD with a
constant accretion rate $\dot{M}_{\rm acc}=6\times10^{-7}\,{M}_\odot\,\mbox{yr}^{-1}$.
The He layer in this case periodically experiences He layer flashes.
The WD quickly enters into a recurring condition of flashes in which mass
is accreted after the first He layer flash. During a He layer flash
some of the envelope matter is lost via the super-Eddington wind
(see also Denissenkov et al. 2013). The positive slope for the  mass growth of the WD shows that the WD grows in
mass at a rate of $6.571\times10^{-8}\,{M}_\odot\,\mbox{yr}^{-1}$ for a constant accretion rate of $6\times10^{-7}\,{M}_\odot\,\mbox{yr}^{-1}$. This represents a mass retention efficiency per cycle of $\approx$11\%.

\begin{figure}[]
\begin{center}
\includegraphics[width=8.5cm,angle=0]{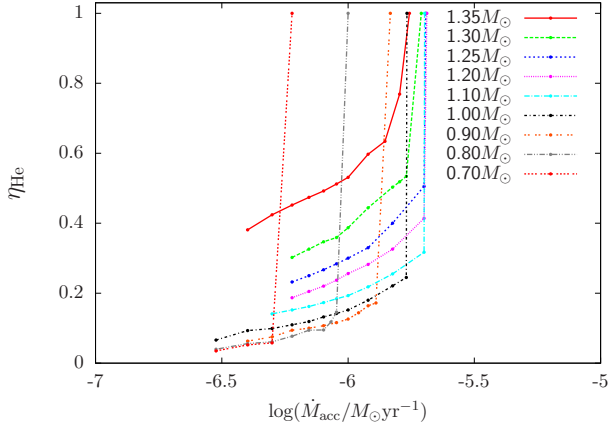}
 \caption{Mass retention efficiency of He layer flashes, $\eta _{\rm He}$, is plotted against
accretion rate for various initial WD masses. }
  \end{center}
\end{figure}

In Fig. 5, we present the mass retention efficiency of He layer flashes for a wide range of
initial WD masses with various accretion rates. The mass retention efficiency $\eta _{\rm He}$ is
a function of the accretion rate and the WD mass. From this figure, we can see that
$\eta _{\rm He}$ increases with the WD mass for a given accretion rate. This is  because
the nuclear burning rates are much higher than the wind mass-loss
rates for more massive WDs due to the strong gravity. Consequently, most of the
layer mass accumulates on the surface of these more massive WDs.
The mass retention efficiency in He layer flashes plays an important  role in  binary
evolution, which has an important influence on
the birthrates of SNe Ia. 
Note that $\eta _{\rm He}$ may be increased if rotation is considered (e.g., Yoon et al. 2014).
The data points in Fig. 5 can be
used in the studies of binary population synthesis computations.
These data points and the corresponding
interpolation FORTRAN code can be provided by contacting the author.

\section{Discussion and conclusions}

When calculating the
Eddington  accretion rate, we did not consider the contribution of $L_{\rm
acc}$ to the total luminosity of the WD  (see also Ma et al. 2013) because the accretion energy  can be
radiated away from the WD surface rapidly (e.g., Townsley \& Bildsten 2004).
If we include ${L}_{\rm acc}$ in  the total luminosity of the WD,
the super-Eddington wind will be formed more
easily as the accretion energy will also contribute to expelling the accreted matter.
Ma et al. (2013) recently investigated the evolution of CO WDs accreting H-rich matter, and
examined the effect of ${L}_{\rm acc}$ on the accretion process of WDs. They found that the accretion luminosity
has no significant influence on the Eddington accretion rate (see Fig. 3 of Ma et al. 2013).
The difference between using $L$ or $L_{\rm prime}$ as a criterion for driving wind
is fairly small, since the nuclear burning luminosity is far greater than the accretion energy.

The super-Eddington He wind may be formed when the luminosity exceeds the Eddington limit,
which can be expected in the observed WD+He star systems.
For WD+He star systems with long orbital periods, the He star will fill its Roche lobe when it
evolves to the subgiant stage (it now contains a CO core).
In this case, the super-Eddington wind would be encountered
owing to a high mass-transfer rate resulting from the rapid
expansion of the He star during the subgiant stage (e.g., Wang et al. 2009a).
HD 49798 (a hydrogen depleted He star) with its massive
WD companion may be a progenitor candidate of SNe Ia.
After about $4\times10^{4}$\,yr, HD 49798 will fill its Roche lobe due to
the rapid expansion of its envelope (see Wang \& Han 2010). In this binary system,
a super-Eddington wind may be triggered on the surface of the massive WD
owing to a high mass-transfer rate soon after the onset of
Roche-lobe overflow (see Fig. 2 in Wang \& Han 2010).

It has been suggested that cataclysmic variables  are progenitor candidates of  SNe Ia
(e.g., Parthasarathy et al.\ 2007; Wang \& Han 2012).
In Fig. 5, we can see that the WD can still undergo mass growth during nova outbursts.
This suggests that some observational cataclysmic variables may be candidates of  SN Ia progenitors.
In the observations, such a He layer
flash could be observed as a He nova outburst.
V445 Puppis was discovered at its outburst stage in late 2000, which
is the first He nova that has been detected so far (e.g., Kato et al. 2000; Ashok \& Banerjee 2003).
Kato et al. (2008) recently fitted
the light curve of V445 Puppis, and argued that the WD mass
is above $1.35\,M_{\odot}$ and half of the accreted
matter is still on the surface of the WD.
Therefore, V445 Puppis is a strong candidate of SN~Ia progenitors (see also Woudt et al. 2009).
However, it is still unclear whether WDs lose more matter during nova outbursts than during the accretion
(e.g., Bours et al. 2013; Idan et al. 2013; Newsham et al. 2014; Hillman et al. 2015).

At low accretion rates (e.g., $<$$\dot{M}_{\rm st}$), the accreting
WD would develop He layer flashes. For a given mass of the CO WD core,
the strength of He layer flashes increases with decreasing
$\dot{M}_{\rm acc}$.
If the accretion rate of the WD is too low (e.g., $<$$4\times10^{-8}\,M_{\odot}\mathrm{yr}^{-1}$),
a thick He layer is believed to grow on its surface. In this case,
a sub-Chandrasekhar double-detonation may happen once a He layer with a critical mass accumulates on the WD
(e.g., Woosley et al.\ 1986; Piersanti et al. 2014).
CD$-$30$^{\circ}$\,11223 has been identified as a
WD+He star system with a $\sim$1.2\,h orbital period (e.g., Vennes
et al. 2012; Geier et al. 2013). The masses of the He star and the WD
are constrained to be $\sim$$0.51\,M_{\rm\odot}$ and $\sim$$0.76\,M_{\rm\odot}$, respectively (see Geier et al. 2013).
Wang et al. (2013b) recently suggested that CD$-$30$^{\circ}$\,11223
might produce an SN Ia through the double-detonation
scenario in its future evolution (see also Geier et al.\ 2013).

Employing the stellar evolution code MESA, we have investigated the
long-term evolution of He accreting WDs.
In order to reproduce the results of the steady-state models, we first
set ${L}_{\rm acc}={L}_{\rm Edd}$ as the wind triggering criteria, and
found that the properties of He burning in our simulations are
almost similar to those of the steady-state models.
If the total luminosity is considered when determining the Eddington accretion rate,
a super-Eddington wind could be formed at relatively lower accretion rates
than those of previous studies of the steady-state models.
The super-Eddington wind can prevent CO WDs with high accretion rates from
becoming red-giant-like He stars.
We also presented the steady He-burning regime in which the He-rich matter can be
burned into carbon and oxygen completely on the surface of the WD. Additionally,
we performed a  multi-cycle evolution of He layer flashes in the simulations of nova outbursts,
and obtained the mass retention efficiency of He layer flashes for various WD masses, which will be useful
in future binary population synthesis studies.
Note also that we used the opacity of the photosphere when we calculate the Eddington
luminosity as was done in Denissenkov et al. (2013).
However, the opacity in the wind may change compared to the values at
the photosphere, which should be further studied.

\begin{acknowledgements}
We acknowledge the anonymous referee for the valuable comments that helped us
to improve the paper. We thank Philipp Podsiadlowski, Pavel A. Denissenkov, Xuefei Chen and Xiangcun Meng 
for their helpful discussions.
We also thank Yan Gao for his kind
help to improve the language of this paper, and thank the computing time granted by the Supercomputing Platform of Yunnan Observatories.
This work is supported by the National Basic Research Program of China (973 programme, 2014CB845700),
the National Natural Science Foundation of China (Nos 11390374, 11521303, 11322327 and U1331117),
the Chinese Academy of Sciences (Nos KJZD-EW-M06-01 and XDB09010202),
and  the Natural Science Foundation of Yunnan Province (Nos 2013HB097 and 2013FB083).
\end{acknowledgements}

\clearpage


\begin{thebibliography}{}
\bibitem[Ashok \& Banerjee (2003)]{ash03}        Ashok, N. M., \& Banerjee, D. P. K. 2003, A\&A, 409, 1007
\bibitem[Badenes et al. (2009)]{bad09}           Badenes, C., Harris, J., Zaritsky, D., \& Prieto, J. L. 2009, ApJ, 700, 727
\bibitem[Bildsten et al. (2007)]{bil07}          Bildsten, L., Shen, K. J., Weinberg, N. N., \& Nelemans, G. 2007, ApJ, 662, L95
\bibitem[Bours et al. (2013)]{bou13}             Bours, M. C. P., Toonen, S., \& Nelemans, G. 2013, A\&A, 552, A24
\bibitem[Brooks et al. (2015)]{bro15}            Brooks, J., Bildsten, L., Marchant, P., \& Paxton, B. 2015, ApJ, 807, 74
\bibitem[Chen et al. (2012)]{che12}              Chen, X., Jeffery, C. S., Zhang, X., \& Han, Z. 2012, ApJL, 755, L9
\bibitem[Chen \& Li (2009)]{che09}               Chen, W.-C., \& Li, X.-D. 2009, ApJ, 702, 686
\bibitem[Claeys et al. (2014)]{cla14}            Claeys, J. S. W., Pols, O. R., Izzard, R. G., Vink, J., \& Verbunt, F. W. M. 2014, A\&A, 563, A83
\bibitem[Denissenkov et al. (2013)]{deni13}      Denissenkov, P. A., Herwig, F., Bildsten, L., \& Paxton, B. 2013, ApJ, 762, 8
\bibitem[Diehl et al. (2014)]{dieh14}            Diehl, R., Siegert, T., Hillebrandt, W., et al. 2014, Science, 345, 1162
\bibitem[Fang \& Zhang (2012)]{fz12}             Fang, J., \& Zhang, L. 2012, MNRAS, 424, 2811
\bibitem[Fink et al. (2007)]{Fin07}              Fink, M., Hillebrandt, W., \& R\"{o}pke, F.~K. 2007, A\&A, 476, 1133
\bibitem[Geier et al. (2015)]{gei15}             Geier, S., F\"{u}rst F., Ziegerer, E., et al. 2015, Science, 347, 1126
\bibitem[Geier et al. (2013)]{gei13}             Geier, S., Marsh, T. R., Wang, B., et al. 2013, A\&A, 554, A54
\bibitem[Geier et al. (2007)]{GEI07}             Geier, S., Nesslinger, S., Heber, U., Przybilla, N., Napiwotzki, R., \& Kudritzki, R.-P. 2007, A\&A, 464, 299
\bibitem[Han \& Podsiadlowski (2004)]{hp04}      Han, Z., \& Podsiadlowski, Ph. 2004, MNRAS, 350, 1301
\bibitem[Han \& Podsiadlowski (2006)]{han06}     Han, Z., \& Podsiadlowski, Ph. 2006, MNRAS, 368, 1095
\bibitem[Hachisu et al. (1996)]{hkn96}           Hachisu, I., Kato, M., \& Nomoto, K. 1996, ApJL, 470, L97
\bibitem[Hachisu et al. (2012)]{hac12}           Hachisu, I., Kato, M., Saio, H., \& Nomoto, K. 2012, ApJ, 744, 69
\bibitem[Hillebrandt et al. (2013)]{hill13}      Hillebrandt, W., Kromer, M., R\"{o}pke, F. K., \& Ruiter, A. J. 2013, FrPhy, 8, 116
\bibitem[Hillman et al. (2015)]{hillm15}         Hillman, Y., Prialnik, D., Kovetz, A., \& Shara, M. M. 2015, submitted to ApJ (arXiv:1508.03141)
\bibitem[Hoeflich et al. (2013)]{hoe13}          Hoeflich, P., Dragulin, P., Mitchell, J., et al. 2013, FrPhy, 8, 144
\bibitem[Howell (2011)]{how11}                   Howell, D. A. 2011, Nature Communications, 2, 350
\bibitem[Iben (1982)]{iben82}                    Iben, I. 1982, ApJ, 259, 244
\bibitem[Iben \& Tutukov (1984)]{it84}           Iben, I., \& Tutukov, A. V. 1984, ApJS, 54, 335
\bibitem[Iben \& Tutukov(1989)]{it89}            Iben, I., \& Tutukov, A. V. 1989, ApJ, 342, 430
\bibitem[Idan et al. (2013)]{ida13}              Idan, I., Shaviv, N. J., \& Shaviv, G. 2013, MNRAS, 433, 2884
\bibitem[Justham (2011)]{jus11}                  Justham, S. 2011, ApJ, 730, L34
\bibitem[Justham et al. (2009)]{Jus09}           Justham, S., Wolf, C., Podsiadlowski, P., \& Han, Z. 2009, A\&A, 493, 1081
\bibitem[Kasliwal et al. (2012)]{kas12}          Kasliwal, M. M., Kulkarni, S. R., Gal-Yam, A., et al. 2012, ApJ, 755, 161
\bibitem[Kato \& Hachisu (2004)]{kat04}          Kato, M., \& Hachisu, I. 2004, ApJL, 613, L129
\bibitem[Kato et al. (2008)]{kh08}               Kato, M., Hachisu, I., Kiyota, S., \& Saio, H. 2008, ApJ, 684, 1366
\bibitem[Kato et al. (2000)]{kat00}              Kato, T., Kanatsu, K., Takamizawa, K., Takao, A., \& Stubbings, R. 2000, IAU Circ, 7552, 1
\bibitem[Langer et al. (2000)]{langer00}         Langer, N., Deutschmann, A., Wellstein, S., \& H\"{o}flich, P. 2000, ApJ, 362, 1046
\bibitem[Li \& van den Heuvel (1997)]{lix97}     Li, X.-D., \& van den Heuvel, E. P. J. 1997, A\&A, 322, L9
\bibitem[Liu et al. (2015)]{liuz15}              Liu, Z.-W., Stancliffe, R. J., Abate, C., \& Wang, B. 2015, ApJ, 808, 138
\bibitem[Livne (1990)]{liv90}                    Livne, E. 1990, ApJ, 354, L53
\bibitem[Ma et al. (2013)]{max13}                Ma, X., Chen, X., Chen, H., Denissenkov, P. A., \& Han, Z. 2013, ApJL, 778, L32
\bibitem[Maoz et al. (2014)]{mao14}              Maoz, D., Mannucci, F.,  \& Nelemans, G. 2014, ARA\&A, 52, 107
\bibitem[Matteucci \& Greggio (1986)]{mat86}     Matteucci, F., \& Greggio, L. 1986, A\&A, 154, 279
\bibitem[McCully et al. (2014)]{mcc14}           McCully, C., Jha, S. W., Foley, R. J., et al. 2014, Nature, 512, 54
\bibitem[Meng \& Podsiadlowski (2013)]{men13}    Meng, X., \& Podsiadlowski, Ph. 2013, ApJL, 778, L35
\bibitem[Meng \& Podsiadlowski (2014)]{men14}    Meng, X., \& Podsiadlowski, Ph. 2014, ApJL, 789, L45
\bibitem[Mereghetti et al. (2009)]{mer09}        Mereghetti, S., Tiengo, A., Esposito, P., et al. 2009, Science, 325, 1222
\bibitem[Nelemans et al. (2001)]{nel01}          Nelemans, G., Portegies Zwart, S. F., Verbunt, F., \& Yungelson, L. R. 2001, A\&A, 368, 939
\bibitem[Newsham et al.  (2014)]{new14}          Newsham, G., Starrfield, S., \& Timmes, F. 2014, ASP Conf. Ser., 490, 287
\bibitem[Nomoto (1982)]{nomoto1982}              Nomoto, K. 1982, ApJ, 253, 798
\bibitem[Nomoto et al. (1984)]{nty84}            Nomoto, K., Thielemann, F., \& Yokoi, K. 1984, ApJ, 286, 644
\bibitem[Parthasarathy et al. (2007)]{par07}     Parthasarathy, M., Branch, D., Jeffery, D. J., \& Baron, E. 2007, New Astron. Rev., 51, 524
\bibitem[Paxton et al. (2011)]{pea11}            Paxton, B., Bildsten, L., Dotter, A., Herwig, F., Lessafre, P., \& Timmes, F. 2011, ApJS, 192, 3
\bibitem[Paxton et al. (2013)]{pea13}            Paxton, B., Cantiello, M., Arras, Ph., et al. 2013, ApJS, 208, 4
\bibitem[Piersanti et al. (2014)]{pier14}        Piersanti, L., Tornamb\'{e}, A., \& Yungelson, L. R. 2014, MNRAS, 445, 3239
\bibitem[Piersanti et al. (2015)]{pier15}        Piersanti, L., Tornamb\'{e}, A., \& Yungelson, L. R. 2015, MNRAS, 452, 2897
\bibitem[Podsiadlowski et al. (2008)]{pod08}     Podsiadlowski, P., Mazzali, P., Lesaffre, P., Han, Z., \& F\"{o}rster, F. 2008, New Astro. Rev., 52, 381
\bibitem[Ruiter et al. (2009)]{rui09}            Ruiter, A. J., Belczynski, K., \& Fryer, C. 2009, ApJ, 699, 2026
\bibitem[Ruiter et al. (2014)]{Rui14}            Ruiter, A. J., Belczynski, K.,  Sim, S. A., Seitenzahl, I. R., Kwiatkowski, D., 2014, MNRAS, 440, L101
\bibitem[Ruiter et al. (2013)]{Rui13}            Ruiter, A. J., Sim, S. A., Pakmor, R., et al., 2013, MNRAS, 429, 1425
\bibitem[Shen \& Bildsten (2007)]{shen07}        Shen, K. J., \& Bildsten, L. 2007, ApJ, 660, 1444
\bibitem[Shen \& Bildsten (2009)]{she09}         Shen, K. J., \& Bildsten, L. 2009, ApJ, 699, 1365
\bibitem[Shen et al. (2010)]{she10}              Shen, K. J., Kasen, D.,Weinberg, N. N., Bildsten, L., \& Scannapieco, E. 2010, ApJ, 715, 767
\bibitem[Sim et al. (2012)]{sim12}               Sim, S. A., Fink, M., Kromer, M., R\"{o}pke, F. K., Ruiter, A. J., \& Hillebrandt, W. 2012, MNRAS, 420, 3003
\bibitem[Tauris et al. (2013)]{tau13}            Tauris, T. M., Sanyal, D., Yoon, S.-C., \& Langer, N. 2013, A\&A, 558, A39
\bibitem[Toonen et al. (2014)]{ton14}            Toonen, S., Claeys, J. S. W., Mennekens, N., \&  Ruiter, A. J.  2014, A\&A, 562, A14
\bibitem[Townsley \& Bildsten (2004)]{tb04}      Townsley, D. M., \& Bildsten, L. 2004, ApJ, 600, 390
\bibitem[Vennes et al. (2012)]{ven}              Vennes, S., Kawka, A., O'Toole, S. J., N$\acute{\rm e}$meth, P, \& Burton, D. 2012, ApJ, 759, L25
\bibitem[Wang \& Han (2009)]{wh09}               Wang, B., \& Han, Z. 2009, A\&A, 508, L27
\bibitem[Wang \& Han (2010)]{wh10}               Wang, B., \& Han, Z. 2010, Res. Astron. Astrophys., 10, 681
\bibitem[Wang \& Han (2012)]{wh12}               Wang, B., \& Han, Z. 2012, New Astron. Rev., 56, 122
\bibitem[Wang et al. (2013b)]{wan13b}            Wang, B., Justham, S., \& Han, Z. 2013b, A\&A, 559, A94
\bibitem[Wang et al. (2010)]{wan10}              Wang, B., Li, X.-D., \& Han, Z. 2010, MNRAS, 401, 2729
\bibitem[Wang et al. (2009a)]{wan09a}            Wang, B., Meng, X., Chen, X., \& Han, Z. 2009a, MNRAS, 395, 847
\bibitem[Wang et al. (2009b)]{wan09b}            Wang, B., Chen, X., Meng, X., \& Han, Z. 2009b, ApJ, 701, 1540
\bibitem[Wang et al. (2014a)]{wan14a}            Wang, B., Meng, X., Liu, D., Liu, Z., \& Han, Z. 2014a, ApJL, 794, L28
\bibitem[Wang et al. (2014b)]{wan14b}            Wang, B., Justham, S., Liu, Z., Zhang, J., Liu, D., \& Han, Z. 2014b, MNRAS, 445, 2340
\bibitem[Wang et al. (2013a)]{wangx13a}          Wang, X.-F., Wang, L., Filippenko, A. V., Zhang, T., \& Zhao, X. 2013a, Science, 340, 170
\bibitem[Webbink (1984)]{web84}                  Webbink, R. F. 1984, ApJ, 277, 355
\bibitem[Whelan \& Iben (1973)]{wi73}            Whelan, J., \& Iben, I. 1973, ApJ, 186, 1007
\bibitem[Woosley \& Kasen (2011)]{woo11}         Woosley, S. E., \& Kasen, D. 2011, ApJ, 734, 38
\bibitem[Woosley et al. (1986)]{woo86}           Woosley, S. E., Taam, R. E., \&  Weaver, T. A. 1986, ApJ, 301, 601
\bibitem[Woudt et al. (2009)]{wou09}             Woudt, P. A., Steeghs, D., Karovska, M., et al. 2009, ApJ, 706, 738
\bibitem[Yoon \& Langer (2003)]{yoo03}           Yoon, S.-C., \& Langer, N. 2003, A\&A, 412, L53
\bibitem[Yoon \& Langer (2004)]{yoo04}           Yoon, S.-C., \& Langer, N. 2004, A\&A, 419, 623
\bibitem[Yoon et al. (2014)]{yoo14}              Yoon, S.-C., Langer, N., \& Scheithauer, S. 2004, A\&A, 425, 217



\end{thebibliography}
\end{document}